# Thermal transport in a 2D amorphous material

Yuxi Wang[1,2,3#], Xingxing Zhang[4#], Wujuan Yan[1,2,3#], Nianjie Liang[1,2,3#], Haiyu He[1,2,3#], Xinwei Tao[4], Ang Li[5], Fuwei Yang[3,6], Buxuan Li[7], Te-Huan Liu[8], Jia Zhu[9], Wu Zhou[5], Wei Wang[2,10], Lin Zhou[4*], Bai Song[1,2,3*]

[1]College of Engineering, Peking University, Beijing 100871, China

[2]National Key Laboratory of Advanced MicroNanoManufacture Technology, Peking University, Beijing, 100871, China

[3]Beijing Innovation Center for Engineering Science and Advanced Technology, Peking University, Beijing 100871, China

[4]School of Chemistry and Chemical Engineering, Frontiers Science Centre for Transformative Molecules, Shanghai Jiao Tong University, Shanghai 200240, China.

[5]School of Physical Sciences and CAS Key Laboratory of Vacuum Physics, University of Chinese Academy of Sciences, Beijing 100049, China.

[6]Center for Nano and Micro Mechanics, Tsinghua University, Beijing 100084, China.

[7]Department of Mechanical Engineering, Massachusetts Institute of Technology, 77 Massachusetts Avenue, Cambridge, MA 02139, USA

[8]School of Energy and Power Engineering, Huazhong University of Science and Technology, Wuhan, Hubei 430074, China

[9]College of Engineering and Applied Sciences, Nanjing University, 210093 Nanjing, China.

[10]School of Integrated Circuits, Peking University, Beijing 100871, China.

[#]These authors contributed equally to this work.

[*]Corresponding author. Email: songbai@pku.edu.cn (B.S.); linzhou@sjtu.edu.cn (L.Z.)

**Abstract**

Two-dimensional (2D) crystals proved revolutionary soon after graphene was discovered in 2004. However, 2D amorphous materials only became accessible in 2020 and remain largely unexplored. In particular, the thermophysical properties of amorphous materials are of great interest upon transition from 3D to 2D. Here, we probe thermal transport in 2D amorphous carbon. A cross-plane thermal conductivity ($\kappa$) down to 0.079 Wm$^{-1}$K$^{-1}$ is measured for van der Waals stacked multilayers at room temperature, which is among the lowest reported to date. Meanwhile, an unexpectedly high in-plane $\kappa$ is obtained for freestanding monolayers which is a few times larger than what is predicted by conventional wisdom for 3D amorphous carbon with similar sp$^2$ fraction. Our molecular dynamics simulations reveal the role of disorder and highlight the impact of dimensionality. Amorphous materials at the 2D limit open up new avenues for understanding and manipulating heat at the atomic scale.

Amorphous materials such as glass have captured human imagination for thousands of years and continue to be of immense value for diverse technologies including advanced manufacturing, high-performance electronics, and thermal barrier coatings, in light of their intriguing structures and fascinating properties[1,2]. In particular, recent years have witnessed substantial progress in terms of fabrication, characterization, and simulation. For example, the atomic structures of glassy solids (e.g., silica and high-entropy alloy) have finally been directly observed[3-5] ever since Zachariasen[6] proposed the continuous-random-network (CRN) model in 1932. Further, extreme mechanical and dielectric properties have been realized in amorphous carbon and boron nitride, respectively[7,8]. Moreover, ab initio and machine-learning-assisted calculations have markedly facilitated the mechanistic understandings of amorphous materials[9-11]. Among all the endeavors, the synthesis of two-dimensional (2D) amorphous carbon in 2020 down to a single atomic layer is especially noteworthy, which offers unprecedented opportunities for achieving deeper insights, unusual properties, and novel applications[12-14]. However, apart from scarce structural, mechanical, and electrical characterizations, there remains a compelling need to explore thermal transport in amorphous materials at the 2D limit[15].

Heat flow in non-metallic crystals are mainly carried by phonons which can lead to ultrahigh thermal conductivities over 2000 Wm$^{-1}$K$^{-1}$ at room temperature (RT), such as in diamond[16,17], owing to the large group velocities and long lifetimes. In contrast, glasses are usually characterized by a low thermal conductivity ($\kappa$) around 1 Wm$^{-1}$K$^{-1}$, as limited by the random walk of vibrational energy among disordered atoms[18,19]. However, notable exceptions exist. For example, unusual high $\kappa_{RT}$ of over 20 Wm$^{-1}$K$^{-1}$ has recently been realized in 3D amorphous carbon with ~95% sp$^3$ bonding[8], which confirms a conventional wisdom that predicts higher $\kappa$ for materials with more sp$^3$ and less sp$^2$ bonding[20,21]. Following this line of thought, the in-plane thermal conductivity of 2D amorphous carbon should reach a minimum due to its 100% sp$^2$ bonding. Furthermore, an ultralow cross-plane $\kappa_{RT}$ down to ~0.04 Wm$^{-1}$K$^{-1}$ has recently been achieved in van der Waals (vdW) stacked 2D crystals of transition metal dichalcogenides (TMDs) with random interlayer orientations[22]. This motivates us to wonder whether the additional intralayer disorder can lead to even lower cross-plane $\kappa$ in stacked 2D amorphous carbon.

In this work, by systematically measuring and simulating thermal transport in 2D amorphous carbon, we attempt to address one important scientific question: how would thermal transport in amorphous materials vary upon transition from 3D to 2D? First, multiple monolayers were assembled into a set of vdW stacks for cross-plane thermal measurement using the laser pump-probe technique of frequency-domain thermoreflectance. Subsequently, for in-plane transport, the microbridge method was employed with the monolayer amorphous carbon (MAC) samples suspended across pairs of custom-fabricated heating and sensing micro-islands. We observed a cross-plane $\kappa_\perp$ down to 0.079 ± 0.012 Wm$^{-1}$K$^{-1}$ at room temperature which is one of the lowest values reported to date, and a remarkable high in-plane $\kappa_\parallel$ up to 5.47 ± 0.32 Wm$^{-1}$K$^{-1}$ contrary to conventional wisdom. In order to understand these unusual observations, we then performed systematic molecular dynamics simulations which highlighted the role of disorder and dimensionality in both directions.

**Basic characteristics of MAC**

Our 2D amorphous carbon samples were grown via atmospheric pressure chemical vapor deposition. For characterizations, the monolayers were delaminated from the growth surface and then transferred to a target substrate. In Fig. 1a, we show a representative optical image of a continuous MAC film on a silica-coated silicon (SiO$_2$/Si) substrate. Using atomic force microscopy (AFM), the intrinsic monolayer thickness was determined to be ~0.4 nm directly on as-grown samples (Fig. 1b and Supplementary Fig. 1), which agrees well with an earlier report[14].

Values of around 0.6 nm were also observed—both by others[13] and in our study (Supplementary Fig. 1), likely due to sample handling and substrate conditions. The slightly larger thickness compared to graphene (also a single layer of carbon atoms) was previously attributed to the out-of-plane buckling due to Stone-Wales defects[13,14]. Further, Raman spectroscopy confirmed the absence of the 2D peak and yielded broad D and G peaks with an intensity ratio of ~0.8 (Fig. 1c), indicating an amorphous carbon network[13,14]. This was corroborated subsequently by selected area electron diffraction. More directly, representative MAC samples were imaged with scanning transmission electron microscopy (STEM), which verified the disordered atomic structure containing randomly distributed pentagons, hexagons, heptagons, and octagons (see Supplementary Information).

**Cross-plane thermal transport**

In order to probe cross-plane thermal transport, monolayer amorphous carbon films were first assembled layer by layer into a vdW stack on a $SiO_2$/Si substrate using two distinct methods—polymer-free and polymer-assisted (Supplementary Note 1 and Supplementary Fig. 2), both with great care to ensure interfacial cleanliness. Subsequently, we employed the approach of frequency-domain thermoreflectance (FDTR) to measure the cross-plane thermal conductance of the stacked MAC samples (see Methods and Supplementary Information), which required the evaporation of a gold (Au) nanofilm as an opto-thermal transducer (Fig. 2i, inset). Here, the sample conductance is obtained as part of the total interfacial thermal conductance ($G$) between the Au film and the $SiO_2$/Si substrate, as usually practiced in laser pump-probe studies of nanometer-thin materials[22,23]. To improve accuracy, key sample parameters such as the thickness and $\kappa$ of the Au film, $\kappa$ of silicon dioxide and silicon, were all independently measured (Supplementary Notes 2-3, Supplementary Figs. 3-4, and Supplementary Table 1), and the signal sensitivities were systematically analyzed (Supplementary Note 4 and Supplementary Fig. 5).

To begin with, a two-layer (2L) stacked MAC sample was prepared by the polymer-free approach, as shown in Fig. 2a. The white box marks a 100 μm × 100 μm area that was selected to perform FDTR mapping of $G$ with a scanning step size of 0.5 μm and data acquisition at six pump-modulation frequencies of high sensitivity for each pixel. The regions labeled 1L and 2L contain mono- and bi-layers, respectively, which match well with the FDTR-measured $G$ map (Fig. 2b). The average $G$ for 1L and 2L MAC were extracted by fitting normal distributions to the histograms in Figs. 2e and f, respectively, which are around 47.8 and 39.6 $MWm^{-2}K^{-1}$, compared to ~100 $MWm^{-2}K^{-1}$ for the bare substrate without MAC.

Subsequently, a set of eight samples including a bare substrate and stacked MAC of 1L, 2L, 6L, 8L, 12L, 14L, and 16L were prepared for more comprehensive investigations. Due to challenges in preparing thick stacks with the polymer-free method, we switched to a polymer-assisted approach that involved mechanical contact with only the top layer (Supplementary Fig. 2), which helped ensure the cleanliness of the internal interfaces. An additional annealing process further helped improve interfacial adhesion (Methods). In Figs. 2c and d, we plot representative $G$ maps with a step size of 0.2 μm for the 8L and 16L samples, respectively, which appear uniform over large areas but can also contain small imperfections. The corresponding histograms of $G$ are shown in Figs. 2g and h, yielding average values around 21.4 and 16.5 $MWm^{-2}K^{-1}$ which are comparable to that of multilayer graphene[23]. In addition to FDTR mapping, we also performed single-spot measurements with fifty modulation frequencies at randomly selected locations within the scanned area. Figure 2i shows the FDTR phase signals for three different samples including the bare substrate, 8L, and 16L MAC, each representing the average of data obtained at 100 different spots and yielding $G$ values consistent with those from $G$ mapping (see Supplementary Table 2).

To obtain the effective cross-plane thermal conductivity ($\kappa_\perp$) of the stacked MAC samples, we plot in Fig. 2j the thermal resistance ($R = 1/G$) from single-spot measurement as a function of film thickness ($t$). Here, $t$ was systematically measured with AFM, yielding an average monolayer thickness of 0.46 nm for the 6L to 16L samples (Supplementary Table 2). For the 1L and 2L samples, higher $G$ were measured after annealing. For the thicker samples, $R$ increases approximately linearly with $t$, although the variations are small and the uncertainties are relatively large. By using $R$ of the 1L sample ($R_{1L}$) as a reference, we can subtract the contributions of the Au/MAC and MAC/SiO$_2$ interfaces from the total interfacial thermal resistance of the thick stacked MAC samples (Supplementary Note 6). The effective cross-plane thermal conductivity is then obtained as $\kappa_\perp = \frac{t-t_{1L}}{R-R_{1L}}$, which is also plotted in Fig. 2j and ranges from 0.079 to 0.112 Wm$^{-1}$K$^{-1}$ for different samples. This is nearly two orders of magnitude lower than the $\kappa_\perp$ of graphite[24] and many other vdW layered crystals, and approaches the lowest $\kappa$ observed among solids[17].

**In-plane thermal transport**

For in-plane thermal transport, we employed the microbridge method (Methods and Supplementary Information) so that freestanding monolayer amorphous carbon could be measured (Fig. 3). To this end, we custom-fabricated a series of microdevices consisting of two closely spaced silicon nitride (SiN$_x$) islands, both of which are suspended by long and narrow SiN$_x$ beams to achieve good thermal isolation from the substrate, and integrated with platinum serpentine resistors for precise heating and temperature sensing (Figs. 3a and b). To measure the in-plane thermal conductivity, a MAC film was first transferred onto the microdevice, and then cut into a well-defined rectangle bridging the heater and sensor islands (Supplementary Fig. 12), as illustrated in Fig. 3b. Subsequently, a sinusoidal electric current ($i$) at a frequency $f$ was supplied to the heater island so its temperature could be elevated to $T_h$ and modulated at $2f$. This generated a heat flow $Q$ at $2f$ through the suspended MAC sample into the sensor island at a lower temperature of $T_s$, which was measured resistively by applying a direct current ($I$). Using this modulated heating scheme, temperature variations and heat currents can be resolved with micro-Kelvin and pico-Watt resolution, respectively (Supplementary Note 8).

In Figs. 3c and d, we show representative SEM images of two microbridge devices together with the suspended MAC samples (MAC-1 and MAC-2). A total of four samples were measured (Supplementary Fig. 13), which represent three independent MAC films, and a range of lengths (2 - 6 μm) and widths (7 - 21 μm), assuming the same thickness of 0.4 nm (Fig. 1b, Supplementary Fig. 1, and Supplementary Table 3). The in-plane thermal conductivity ($\kappa_\parallel$) of all four samples were obtained in the temperature range from 80 K to 450 K under high vacuum. As shown in Fig. 3e, $\kappa_\parallel$ consistently increases with increasing temperature, which is one of the most prominent features of thermal transport in 3D amorphous materials, as evidenced by the literature data for tetrahedrally-bonded amorphous carbon (ta-C) and hydrogenated ta-C (ta-C:H)[25,26]. At room temperature, the $\kappa_\parallel$ values of these single-atomic-layer amorphous carbon films are around 4.49 - 5.47 Wm$^{-1}$K$^{-1}$, which are about three orders of magnitude lower than that of graphene[27], but generally higher than those of 3D amorphous carbon. Variations (~20%) of $\kappa_\parallel$ across different samples are noticed, although a quantitative and microscopic understanding is not yet available. Interestingly, the normalized $\kappa_\parallel$ values with respect to $\kappa_{RT}$ follow very similar trend (Fig. 3e, inset). These observations imply that the in-plane thermal transport is dominated by the disordered structure, while the exact atomic arrangement and inevitable imperfections in sample preparation can also have non-trivial effects.

**Molecular dynamics simulations**

In addition to experimental measurements, we performed systematic molecular dynamics (MD) simulations and incorporated quantum corrections (Methods and Supplementary Information). Atomic models of monolayer amorphous carbon with dimensions up to 30 nm × 30 nm were constructed using the kinetic Monte Carlo method, which began with randomly distributed carbon atoms in 2D and proceeded to lowering the system energy via probabilistic Stone-Wales transformations (Supplementary Fig. 17). Comprehensive structural analyses were conducted in terms of the ring size, bond length, bond angle, and the pair correlation function (Supplementary Fig. 18), yielding results consistent with our STEM observations and previous reports[13,14]. Schematic illustrations of the non-equilibrium simulations for the cross-plane and in-plane transport are presented in Supplementary Fig. 19, which show the atomic structures and directions of heat flow.

In Fig. 4a, we plot the cross-plane and in-plane thermal conductivity for stacked MAC and freestanding monolayers, respectively, as a function of temperature. The calculated $\kappa_{\perp,RT}$ of stacked MAC is around 0.05 Wm$^{-1}$K$^{-1}$, lower than the experimental values of ~0.08 Wm$^{-1}$K$^{-1}$. The calculated $\kappa_\parallel$ increases slightly with model dimensions and ranges from 5.87 to 7.35 Wm$^{-1}$K$^{-1}$. Despite quantitative discrepancies, an overall agreement between the simulations and experiments is observed (Supplementary Fig. 16a). Moreover, upon normalization, all the measured and calculated $\kappa_{\parallel,RT}$ overlap well, in terms of the temperature dependence (Supplementary Fig. 16b). These results suggest that our simulations have captured the essential mechanisms of thermal transport in MAC. For comparison, thermal transport in randomly stacked graphene and 3D amorphous carbon dominated by sp$^2$ bonding (~94%) were also simulated (Supplementary Fig. 19), using the same interatomic potentials. Both feature a thermal conductivity that increases with temperature, as is characteristic of amorphous materials. The calculated $\kappa_\perp$ of the former closely follows that of stacked MAC (Fig. 4a), while the predicted $\kappa$ of the latter (3.20 Wm$^{-1}$K$^{-1}$) is notably lower than the $\kappa_\parallel$ of freestanding MAC.

**Role of disorder and dimensionality**

Carbon forms a rich variety of allotropes and almost limitless compounds, owing to its capacity for highly diversified chemical bonding[28]. In Figs. 4a and b, we summarize previously reported thermal conductivities of representative carbon allotropes from 0D to 3D covering both crystalline and glassy materials. These include diamond[16,29], graphite[24,30], graphene[27,31], single-walled carbon nanotube (SWCNT)[32], fullerene[33,34], 3D amorphous carbon[8,20,21,25,26], and stacked graphene[23]. Intriguingly, $\kappa_{RT}$ of all carbon allotropes spans five orders of magnitude from ultrahigh (>1000 Wm$^{-1}$K$^{-1}$) to ultralow (~0.01 Wm$^{-1}$K$^{-1}$), which is basically the full range for thermal transport in solids. In particular, even amorphous carbon alone can have $\kappa_{RT}$ values varying by 1000-fold. These characteristics reflect the great breadth and depth of the underlying physical mechanisms, and bring us back to the scientific question that motivated this work, namely, the role of disorder combined with dimensionality.

First, in the cross-plane direction, our measured and calculated thermal conductivity for stacked MAC are around 0.08 Wm$^{-1}$K$^{-1}$ and 0.05 Wm$^{-1}$K$^{-1}$ (Fig. 4a), respectively, which are among the lowest values ever reported[22,35,36]. This ultralow thermal conductivity originates from the synergistic interplay between the cross-plane structural disorder and the weak interlayer vdW interaction characteristic of 2D materials. The comparable $\kappa_{\perp,RT}$ of stacked MAC and graphene is rather counterintuitive considering the additional in-plane disorder of the former, however, an intuitive physical picture can be proposed. Briefly, both stacks feature very small interlayer shear moduli due to their random and thus incommensurate interfaces, which underlie the structural superlubricity in graphite[37] and lead to substantial softening of the transverse vibrational modes.

In addition, the longitudinal modes are strongly affected by the cross-plane disorder which is shared by stacked MAC and graphene. Together, the similarities in both the transverse and longitudinal modes (Supplementary Fig. 20) explain the similar cross-plane conductivities.

Next, we consider the role of dimensionality for in-plane transport. It is known that dimensionality can impact a plethora of phenomena such as wave localization and hydrodynamic flow[38]. Also, substantial enhancement of thermal conductivity with reducing dimensionality has been reported[39,40]. In Fig. 4b, we illustrate dimensional crossover in the thermal conductivity of both crystalline and amorphous carbon, as the atomic coordination transforms from pure $sp^3$ to $sp^2$. Notably, we observe for the first time that the $\kappa_\parallel$ values of 2D MAC (~100% $sp^2$) are discernibly higher than those of 3D amorphous carbon with similarly high percentage of $sp^2$ bonding (Fig. 4b). This is especially unusual considering the traditional expectation that for 3D amorphous carbon, more $sp^2$ means lower thermal conductivity[8,20,21].

In order to gain more physical insights, we calculated the dispersion relations for the vibrational modes in both 3D amorphous carbon and MAC (Figs. 4c-f and Supplementary Fig. 21)[19]. For 3D amorphous carbon with 94% $sp^2$ bonding, the dispersions for both the longitudinal and transverse modes are almost entirely diffuse (Figs. 4c and d). In comparison, despite the complete disorder and pure $sp^2$ bonding, the dispersions for MAC clearly feature well-defined branches similar to those of graphene even at relatively high frequencies, for example, up to ~20 THz for the longitudinal modes (Figs. 4e and f). This suggests that low-frequency propagating waves contribute dominantly to the unusual high $\kappa_\parallel$ of MAC. The crucial role of the low-frequency modes in MAC is further corroborated by their notably higher vibrational density of states (VDOS) and participation ratio, as shown respectively in Figs. 4g and h. Ultimately, these distinct characteristics can be traced back to the 2D nature of MAC which significantly promotes out-of-plane vibrations. As an example, we compare representative eigenvectors for 3D amorphous carbon and MAC as visualized in Figs. 4i and j, respectively. For the former, a localized mode at 2.34 THz is identified, as characterized by an overall random distribution of the vibrational directions and only a few atoms with large amplitudes. In contrast, a propagating mode at 2.39 THz is shown for MAC, as manifested by the consistent out-of-plane atomic vibrations together with more evenly distributed amplitudes.

**Conclusions**

In summary, we have grown 2D amorphous carbon films and revealed thermal transport properties distinct from 3D amorphous carbon, both along the in-plane and cross-plane direction. The ultralow cross-plane and unexpectedly high in-plane thermal conductivity arising from the unique atomic structures and bonding anisotropy in (stacked) MAC highlight 2D amorphous materials as a rich yet unchartered territory for pushing the limit of thermal transport. For the foreseeable future, it would be of interest to quantitatively correlate thermal transport with detailed topological and structural characteristics, and to explore the possibility of directly observing the impact of localization. Finally, we expect the unusual thermal properties of 2D MAC combined with its chemical stability, mechanical strength, and electrical tunability to be uniquely beneficial to various thermal management and energy harvesting applications.

## Methods

**Microscopic characterizations**

Systematic microscopic characterizations of amorphous carbon films were performed using a variety of techniques, with representative data presented in Figs. 1-3 and Supplementary Figs. 1, 3, 12, and 13. The optical images were taken with Olympus BX53. The thickness of monolayer 2D amorphous carbon (MAC) was measured using atomic force microscopy (AFM, MFP-3D Infinity from Asylum Research and Multimode 8 from Bruker). The Raman spectra of MAC were measured using LabRAM HR Evolution, HORIBA Scientific. The scanning electron microscopy (SEM) images were obtained by Helios G4 UX, while selected area electron diffraction was performed with Talos F200X, both from Thermo Fisher Scientific. The scanning transmission electron microscopy (STEM) images were acquired with NION HERMES-100.

**Frequency-domain thermoreflectance**

The cross-plane thermal transport was measured by the widely used technique of frequency-domain thermoreflectance (FDTR). Our FDTR system is based on two continuous-wave (CW) lasers[16,41-43]. Briefly, a 405 nm pump beam is modulated to periodically heat the sample surface and a 532 nm laser probes the surface temperature change at the modulation frequency based on variation of the reflectance. The effective laser spot size (root-mean-square average of the pump and probe beam radii) is 3.3 μm with a 10× objective. A balanced photodetector is used together with a lock-in amplifier to enhance the signal-to-noise ratio. The phase signal was fitted to a Fourier conduction model to obtain the thermal properties of the sample. Both FDTR mapping and single-spot measurement were performed, with six and fifty sampling frequencies, respectively.

**Suspended microbridge**

The in-plane thermal transport was measured using the suspended microbridge method[44,45]. To this end, we first custom-designed and fabricated a series of microbridge devices at Peking University. The device consists of two silicon nitride ($SiN_x$) islands which are suspended and thermally isolated from the substrate by long and narrow $SiN_x$ beams, and also integrated with thin film serpentine platinum resistors (Figs. 3a and b) for precise heating and temperature sensing. In a typical experiment, one of the $SiN_x$ islands is used as the heater, while the other one serves as the sensor. The sample acts as a bridge for heat flow between the heater and the sensor, the thermal conductivity of which can be obtained as $\kappa = G_{MAC}\frac{L}{A} = \frac{Q}{T_h - T_s}\frac{L}{A}$. Here, $G_{MAC}$ is the sample thermal conductance. The temperatures of the two islands ($T_h$ and $T_s$, respectively) are readily monitored with the platinum resistors. The heat current through the sample ($Q$) is measured based on $T_s$ and the beam conductance $G_b$ (typically ~330 nWK$^{-1}$) of the sensor island. The sample length $L$ and cross-sectional area $A$ are obtained using SEM and AFM.

**Construction of atomic model**

The atomic structure of MAC (Supplementary Note 11 and Fig. 17) was generated by the kinetic Monte Carlo (kMC) method[46]. Models with 60% hexagonal rings were chosen as representatives for subsequent structural characterizations and thermal simulations. The distributions of ring size, bond length, and bond angle, together with the pair correlation function of the MAC models (Supplementary Fig. 18a-d) were analyzed through the Atomes software (https://atomes.ipcms.fr/)[47]. For stacked MAC, 16 different MAC structures with a common lateral size were first generated by kMC. Then, stacked MAC containing 8L, 10L, 12L, 14L, and 16L were formed by vertically assembling these monolayers (Supplementary Note 12). Further, randomly-stacked graphene with the same number of layers were created by employing an algorithm that was previously used to generate twisted bilayer graphene[22,48] (Supplementary Note 12). The models of 3D amorphous carbon were generated through a melt-quench method[21] (Supplementary Note 13). All atomic structures were visualized through the OVITO software[49].

**Molecular dynamics simulations of thermal transport**

Simulations of thermal transport in freestanding and stacked MAC, stacked graphene, and 3D amorphous carbon (Supplementary Note 14 and Fig. 19) were performed with the Large-scale Atomic/Molecular Massively Parallel Simulator (LAMMPS) package using nonequilibrium molecular dynamics[50,51]. For freestanding MAC and 3D amorphous carbon, the optimized Tersoff potential was used to describe the interatomic interactions[52,53]. For cross-plane thermal transport in stacked MAC and graphene, the interlayer interaction was described by the Lenard-Jones potential. The vibrational density of states (VDOS) was obtained by Fourier transforming the velocity autocorrelation function. The participation ratio can then be calculated based on the local VDOS[54].

To obtain the dispersion relations, the MD trajectories of the simulated structures in this study were used to analyze the longitudinal and transverse current correlation function through the DYNASOR code[55]. We also calculated the phonon dispersion of graphene using lattice dynamics via the PHONOPY package[56], yielding results consistent with MD simulations. Representative eigenvectors of 3D amorphous carbon and MAC were obtained using PHONOPY.

**Data availability:** All data needed to evaluate the conclusions in the paper are present in the main text or the supplementary information.

**Method references**

**Acknowledgments**

We thank the Molecular Materials and Nanofabrication Laboratory, Electron Microscopy Laboratory, and the High-performance Computing Platform of Peking University for supporting our experiments and simulations, respectively. B.S. appreciates Dr. Alexei A. Maznev at the Massachusetts Institute of Technology for many helpful discussions. This work was financially supported by the National Natural Science Foundation of China (Nos. 52076002, 52103344, and 52076089), Ministry of Science and Technology of China (No. 2022YFA1203100), National Key Basic R&D Program of China (No.2021YFA1401400), and Beijing Outstanding Young Scientist Program (No. BJJWZYJH01201914430039). B.S. acknowledges support from the New Cornerstone Science Foundation through the XPLORER PRIZE.

**Author contributions**

B.S. and L.Z. conceived and supervised the research. X.Z. and X.T. grew the sample. Y.W., W.Y., H.H., and F.Y. conducted the thermal measurements and analyses under the close guidance of B.S. N.L. performed the atomistic simulations. Y.W. and H.H. fabricated the microbridge devices. Y.W., W.Y., H.H., F.Y., X.Z., and X.T. performed the optical, AFM, Raman, and electron diffraction characterizations. A.L. and W.Z. performed the STEM imaging and analysis. W.W., J.Z., T.-H.L., and B.L. provided technical supports. The manuscript was prepared by B.S. and Y.W. with input from all co-authors.

**Competing interests**

Authors declare no competing interests.

**Additional information**

**Supplementary information is available for this paper.**

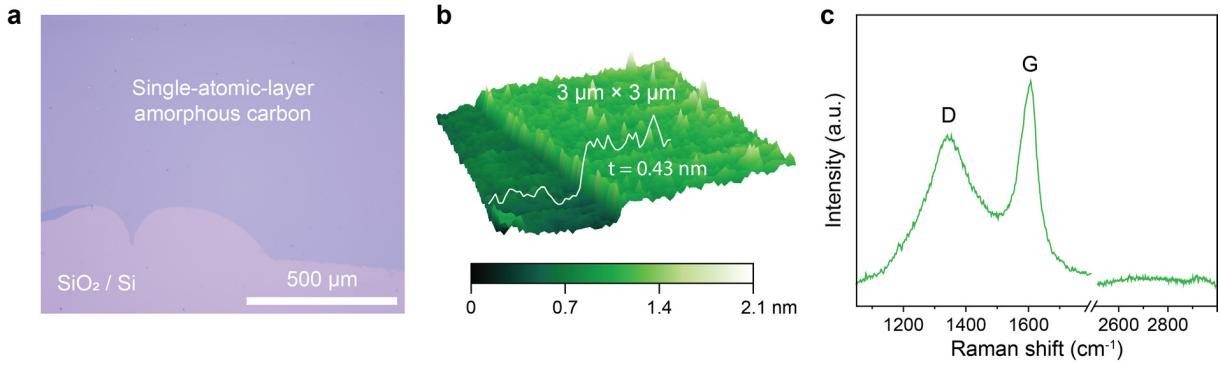

**Fig. 1 | Microscopic and spectroscopic characterizations of 2D amorphous carbon. a,** Optical image of single-atomic-layer amorphous carbon transferred onto a SiO$_2$/Si substrate. **b,** Atomic force microscopy mapping of an as-grown monolayer amorphous carbon sample across the edge. The line profile shows the typical thickness. **c,** Representative Raman spectrum of monolayer amorphous carbon.

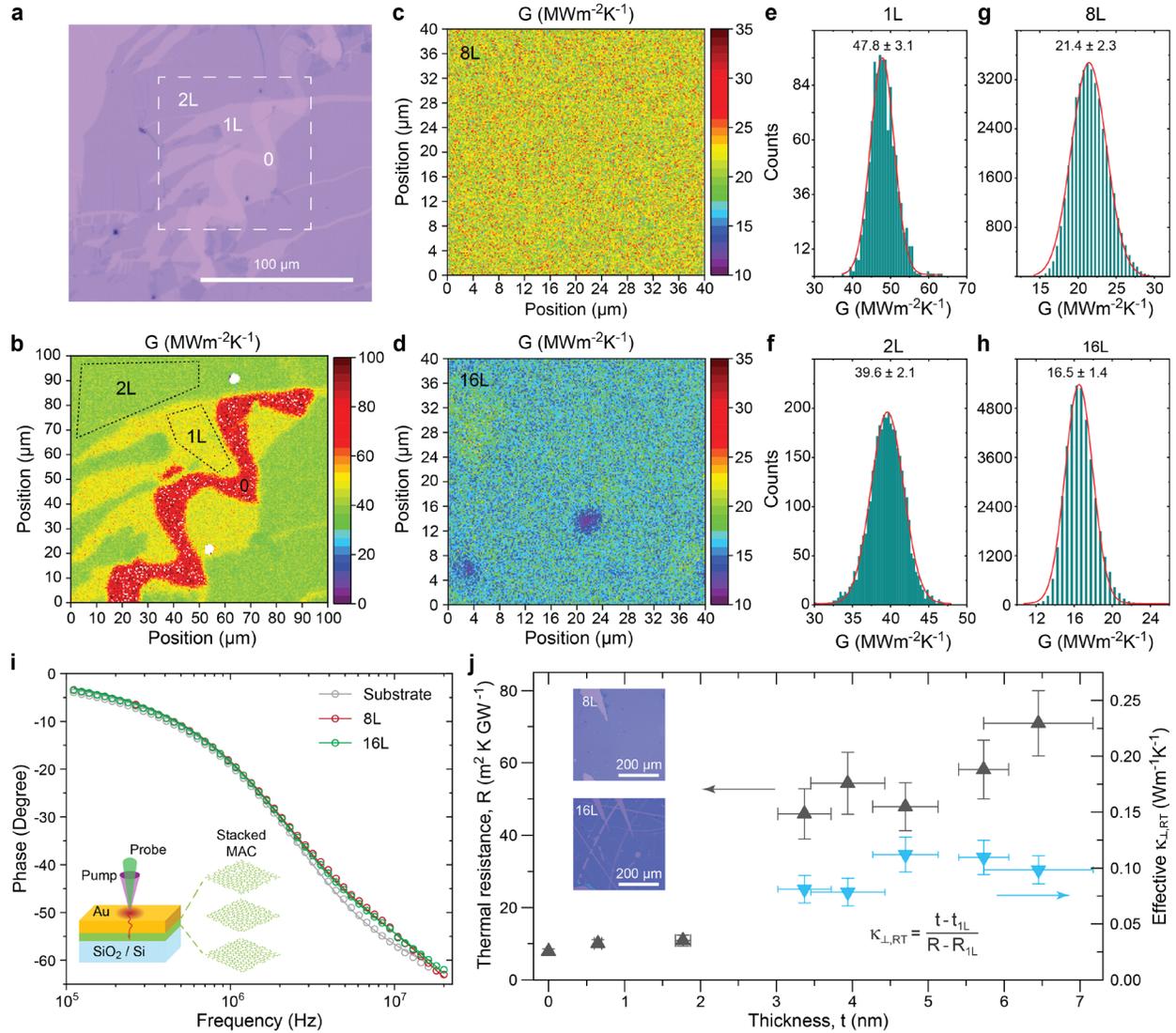

**Fig. 2 | Cross-plane thermal transport measurement in stacked 2D amorphous carbon. a,** Optical image showing 1L and 2L regions on a SiO$_2$/Si substrate (0). **b,** FDTR mapping of the interfacial thermal conductance $G$ within the white square in (**a**). $G$ map of (**c**) 8L and (**d**) 16L MAC. Histograms of $G$ for the (**e**) 1L and (**f**) 2L areas marked in (**b**). Histograms of $G$ for (**g**) 8L and (**h**) 16L MAC obtained from (**c**) and (**d**). **i,** FDTR signals (circles) and fittings (lines) as a function of modulation frequency measured on 8L and 16L samples and also the bare substrate. Inset shows a schematic of the experiment. **j,** Thermal resistance of stacked MAC as a function of thickness for 1L, 2L, 6L, 8L, 12L, 14L, and 16L samples. Data for the substrate is also included. The effective cross-plane thermal conductivity ($k_{\perp,RT}$) for thick samples ($\geq$ 6L) is also plotted, with $t_{1L}$ = 0.46 nm. The error bars include both the experimental noise and uncertainties from various input parameters for the fitting. Inset shows optical images of 8L and 16L samples.

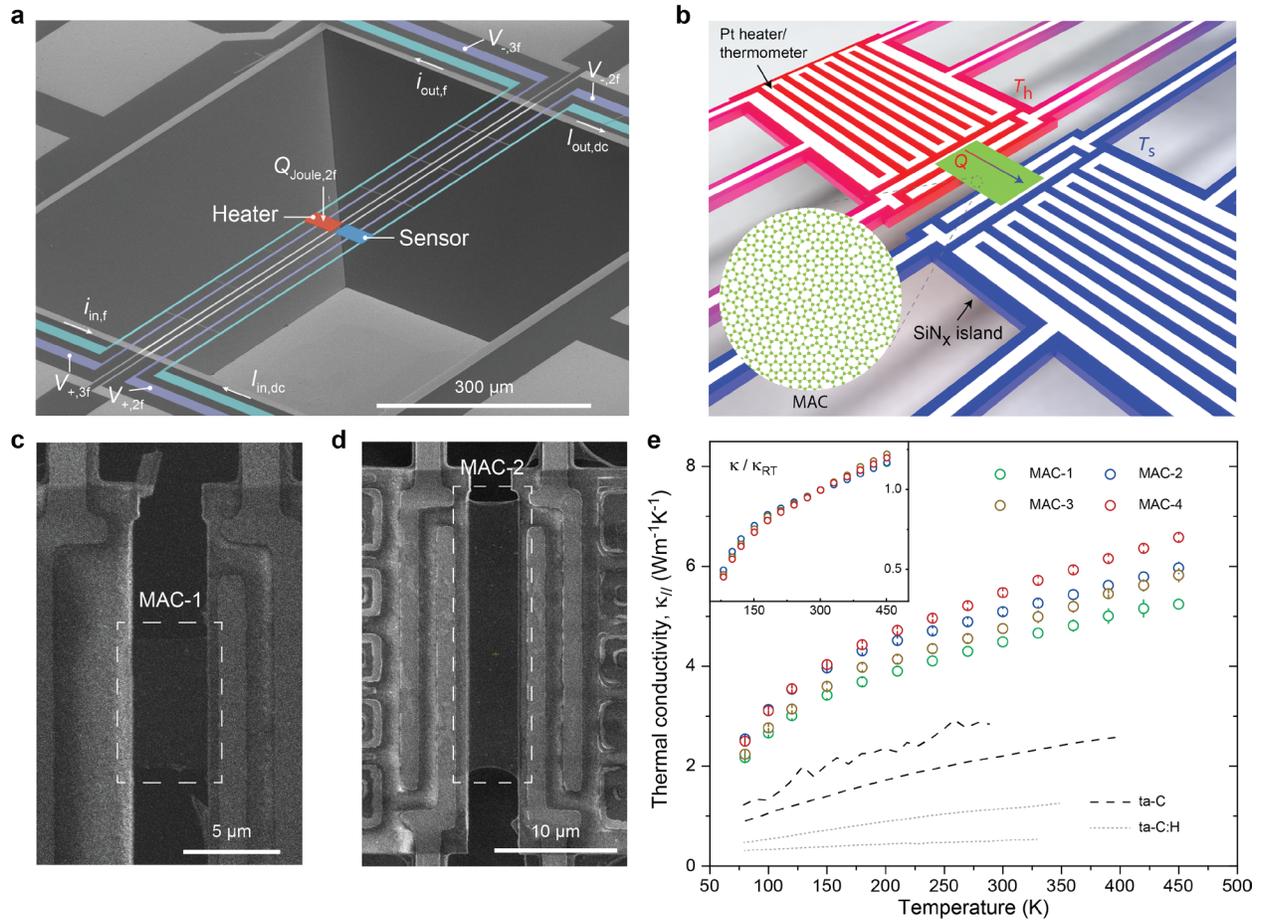

**Fig. 3 | In-plane thermal transport measurement of freestanding 2D amorphous carbon. a,** False-colored scanning electron microscopy (SEM) image of a suspended microbridge device with the key electrical and thermal signals labelled. **b,** Schematic of a MAC sample bridged on a microbridge device. Inset shows the simulated atomic structure. **c** and **d,** SEM images of microbridge devices with suspended MAC samples. **e,** Measured in-plane thermal conductivities of four MAC samples (1 and 2 are from the same batch) as a function of temperature, together with literature data for thin films of 3D amorphous carbon[25,26]. Inset shows the normalized thermal conductivity of MAC with respect to the room temperature values. The error bars represent uncertainties in the thermal conductance measurement.

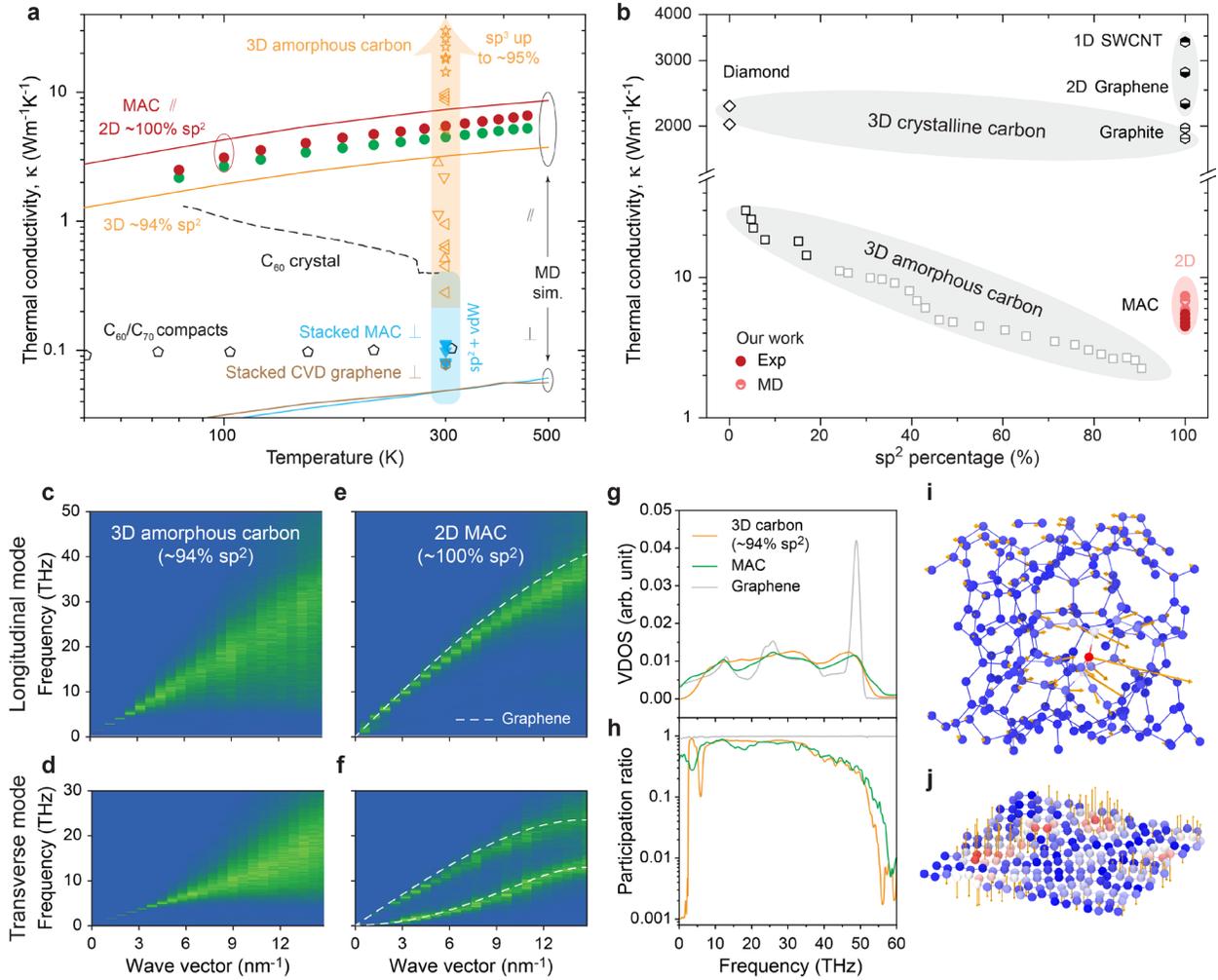

**Fig. 4 | Molecular dynamics simulation of thermal transport in 2D amorphous carbon. a,** Calculated thermal conductivities of (stacked) MAC plotted against the experimental values, together with literature data for a selection of carbon allotropes with low $k$, including fullerene[33,34], 3D amorphous carbon[8,20,25,26], and stacked polycrystalline graphene[23]. The thermal conductivities of 3D amorphous carbon with 94% $sp^2$ bonding and randomly stacked graphene are also calculated and presented for comparison. **b,** Thermal conductivity of crystalline and amorphous carbon allotropes, including diamond[16,29], graphite[24,30], graphene[27,31], single-walled carbon nanotube[32], and 3D amorphous carbon[8,21], as a function of $sp^2$ percentage. **c-f,** Dispersion relations of the longitudinal and transverse vibrational modes for 3D amorphous carbon and MAC along the Γ-M direction. The phonon dispersion for graphene is calculated using lattice dynamics and plotted for comparison. **g** and **h,** The vibrational density of states and participation ratio, respectively, for 3D amorphous carbon, MAC, and graphene. **i** and **j,** Visualization of representative eigenvectors for 3D amorphous carbon (2.34 THz) and MAC (2.39 THz), respectively.

23